\documentclass[fleqn,usenatbib]{mnras}
\usepackage{amsmath}
\usepackage{graphicx}
\usepackage{grffile}
\usepackage{upgreek}

\PassOptionsToPackage{pdfpagelabels=true}{hyperref}

\usepackage{txfonts}
\usepackage{natbib}
\usepackage{bm}
\usepackage{calc}
\usepackage{color}

\usepackage[T1]{fontenc}
\usepackage{ae,aecompl}

\usepackage[final,xcolor={divpdf,red}]{changes}

\title[The post-maximum behaviour of NGC~1566] {The  post-maximum behaviour of the changing-look Seyfert galaxy NGC~1566}

\author[V. L. Oknyansky et al.]{V. L. Oknyansky,$^{1}$\thanks{E-mail: oknyan@mail.ru}
H. Winkler,$^{2}$
S. S. Tsygankov,$^{3,4}$
V. M. Lipunov,$^{1,5}$
\newauthor E. S. Gorbovskoy,$^{1}$
F. van Wyk,$^{2}$
D. A. H. Buckley,$^{6}$
B.W. Jiang,$^{7,8}$
\newauthor N. V. Tyurina$^{1}$
\\
\\
$^{1}$ Sternberg Astronomical Institute,M. V. Lomonosov Moscow State University,  119234, Moscow, Universitetsky pr-t, 13, Russia\\
$^{2}$ Dept. Physics, University of Johannesburg, PO Box 524, 2006 Auckland Park, South Africa\\
$^{3}$ Department of Physics and Astronomy, FI-20014 University of Turku, Finland \\
$^{4}$ Space Research Institute of the Russian Academy of Sciences, Profsoyuznaya Str. 84/32, Moscow 117997, Russia \\
$^{5}$ Faculty of Physics, Moscow M.V. Lomonosov State University, Leninskie gory 1, Moscow, 119991, Russia\\
$^{6}$ The South African Astronomical Observatory, PO Box 9, 7935 Observatory, South Africa\\
$^{7}$ Key Laboratory for Particle Astrophysics, Institute of High Energy Physics, Chinese Academy of Sciences, \\ 19B Yuquan Road, Beijing 100049, China\\
$^{8}$ School of Astronomy and Space Science, University of Chinese Academy of Sciences, \\ 19A Yuquan Road, Beijing 100049, China
}

\date{Accepted XXX. Received YYY; in original form ZZZ}

\pubyear{2020}
\begin{document}
\date{Received ... Accepted ...}
\pagerange{\pageref{firstpage}--\pageref{lastpage}}
\maketitle{}

\label{firstpage}

\begin{abstract}
 We  present results of the long-term multi-wavelength study of optical, UV and X-ray variability of the nearby changing-look  (CL) Seyfert NGC~1566 observed with the {\it Swift} Observatory and the MASTER Global Robotic Network from 2007 to 2019. We started spectral observations with South African Astronomical Observatory 1.9-m telescope soon after the brightening was discovered in July 2018 and present here the data for the interval between Aug. 2018  to Sep. 2019. This paper concentrates on the remarkable post-maximum behaviour after July 2018 when all bands decreased with some fluctuations. We observed three significant re-brightenings in the post-maximum period  during 17 Nov. 2018--10 Jan. 2019, 29 Apr.--19 Jun. 2019 and 27~ Jul.--6 Aug. 2019. An X-ray flux minimum occurred in  Mar. 2019.  The UV minimum occurred about 3 months later.  It was accompanied by a decrease of the $L_{\rm uv}/L_{\rm x}$ ratio.  New post-maximum spectra covering (31 Nov. 2018 -- 23 Sep. 2019) show dramatic changes compared to 2 Aug. 2018, with fading of the broad lines and [Fe X] $\lambda$6374 until Mar. 2019.  These lines became somewhat brighter in Aug.-Sep. 2019. Effectively, two CL states were observed for this object: changing to type 1.2 and then returning to the low state  as a type 1.8~Sy. We suggest that the changes are due mostly to fluctuations in the energy generation. The estimated Eddington ratios are about 0.055$\%$ for minimum in 2014 and  2.8$\% $ for maximum in 2018.
\end{abstract}




\section{Introduction}

Intensive study of the southern hemisphere active galactic nucleus (AGN) NGC~1566 began in the 1960s shortly after the discovery of quasars \citep{Silva2017, Oknyansky2019, Oknyansky2019b, Parker2019}. NGC 1566 is a nearby face-on Seyfert galaxy \citep{Vaucouleurs1961, Vaucouleurs1973, Shobbrook1966} and the nearest ``changing-look" (CL) AGN (\citealt{Oknyansky2018b}, \citealt{Oknyansky2019} = Paper I).  For details of the definition of CL AGNs, statistics and references see \cite{Shappee2014, MacLeod2019, Ruan2019,  Runnoe2016}. Because of its proximity ($D\approx7.2$ Mpc, see comments in Paper I) NGC~1566 is one of the best objects for studying the CL phenomenon. It was one of the first known AGNs with spectral and photometric variability.  Variability was discovered in 1969 by \cite{Pastoriza1970} when the broad H$\upbeta$ line was found to be significantly weaker than in earlier spectra from 1956 \citep{Vaucouleurs1961} and 1962 \citep{Shobbrook1966}. This was shortly after the discovery of variability of Seyfert galaxies  \citep{Fitch1967}.  During next two decades, NGC~1566 was intensively monitored \citep{Quintana1975, Alloin1985, Winkler1992, Baribaud1992}.
\cite{Quintana1975} published the first optical light curve, covering the period of 1955-1971.  This correlated with the spectral variations, and they noted that NGC~1566 became a ``weak Seyfert" after 1969. The object was in such low state with very weak broad permitted lines in its optical spectrum for about next  12 years \citep{Alloin1986}. Occasional brightenings of H$\upbeta$ were noted in the period 1982-1991 \citep{Alloin1985, Alloin1986, Kriss1991, Winkler1992, Baribaud1992}, but these surges were not as strong as the outbursts of 1962 and 2018. The first photoelectric {\it UBV} monitoring of the object was obtained in 1974-1975 by \citet{Penfold1979}, who also collated all photometric data for 1954-1975 and confirmed that broad H$\upbeta$ intensity variations correlated with optical variability. It is interesting that after the maximum of 1962 NGC~1566 had a deep minimum and then brightened again on a short time scale (months) in 1963. Such re-brightenings soon after deep minima might be a common feature of the variability of NGC~1566 and other CL AGNs.  Also, if an AGN is known to be a CL AGN, then it may show  dramatic variability in the future \citep{MacLeod2019}.   For NGC~1566 at least two dramatic brightenings with changing looks have been observed, one in 1962 and the other in 2018. NGC~1566 was not called a ``Changing Look" AGN before 2018 \citep{Oknyansky2018b} because this terminology only came into common use  over the last few years.

 The most intensive and longest duration previous photometric monitoring only covered the IR \citep[\textit{JHK};][]{Glass2004}. Bright states in the IR were observed in 1982 and 2000. In 1982 the stronger broad line emission was clearly seen, but no spectra were obtained in 2000. The gaps in spectral coverage were sometimes too long, meaning that possible CL episodes with dramatic spectral changes could well have been missed. It is entirely possible that high states of NGC~1566 such as in 1962 \citep{Shobbrook1966, Pastoriza1970} and 2018 (see \citealt{Oknyansky2018b} and Paper I) are recurrent events happening on time-scales of several decades. Smaller amplitude recurrent variations on slightly shorter time-scales of $<10$ years have also been suspected before \citep{Alloin1986}.

The first multi-wavelength investigations of the variability of NGC~1566  (X-ray, UV, optical and IR), as well as the first IR reverberation mapping and investigation of variability of UV/optical emission lines, were published by \cite{Baribaud1992}. It is probable that the time delays ($\sim$ several months) they found are significantly overestimated (presumably because of the low cadence of the observations) and that more realistic ones are probably less than 20 days \citep{Oknyansky2001}.  The size of the broad-line region (BLR) is typically at least a few times smaller than the distance from the central source to the inner edge of the dust torus \citep[see e.g., ][]{Netzer2015} and so the expected time delays in variability of broad emission lines might be just a few days.

Most subsequent studies of NGC~1566 published over the past three decades have not been investigations of spectral and photometric variability but rather of the morphology and properties of the circumnuclear environment \citep[see e.g., ][]{Silva2017, Combes2019} and on large scales \citep{Elagali2019}. Only recently, after the discovery of the strong outburst in the X-ray, UV, optical and IR-continua peaking in July 2018 \citep[see details and references in ][]{Oknyansky2019, Parker2019} has intensive  multi-wavelength and spectral monitoring been initiated and the changing look in the optical spectrum revealed (Paper I and then \citealt{Kollatschny2020}).  

The spectral transition during the CL event of NGC~1566 not only manifested in dramatic intensity changes of the broad emission lines but also in strengthening of high-ionisation coronal lines such as [\ion{Fe}{x}] $\lambda$6374. After maximum was reached in July 2018, fluxes in all bands declined, with some  re-brightenings in December 2018 \citep{Grupe2018b} and at the end of May 2019 \citep{Grupe2019}. Such post-maximum re-brightening episodes are probably typical for some fraction of CL AGNs since such events have been noted in several other CL AGNs  \citep{Oknyansky2017a, Oknyansky2018b, Katebi2018}.

In this paper we present results of the continuation of our multi-wavelength (optical, UV and X-ray) monitoring of NGC~1566 using the data obtained with {\it Neil Gehrels Swift Observatory} and the MASTER Global Robotic Network over the period 2007-2019 initiated by \cite{Oknyansky2019}. Preliminary results have appeared also in \cite{Oknyansky2019b}. Here we provide more complete data and results and additional spectral,  photometric (optical, UV) and X-ray data collected up to December 2019. We investigate in detail the post-maximum  behaviour of the object and report an additional re-brightening in August 2019. We also present the analysis of our optical spectra obtained with the South African Astronomical Observatory 1.9-m telescope from Aug. 2018 to Sep. 2019.

\section{Observations, instruments and data reduction}

\subsection{{\it Swift}: X-ray , Ultraviolet, and Optical Observations}

\begin{figure}
\includegraphics[scale=0.8,angle=0,trim=0 0 0 0]{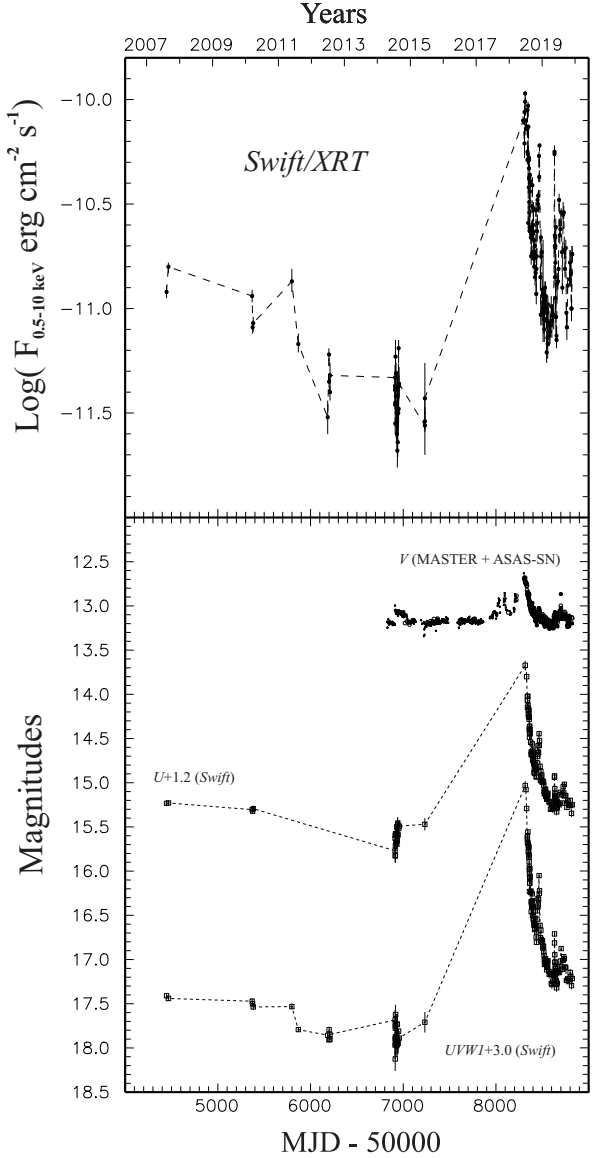}
 \caption {Multi-wavelength observations of NGC~1566 spanning the period 2007 Dec. 11 to 2019 Dec. 5. {\it Top panel:} The {\it Swift}/XRT 0.5--10 keV  X-ray flux (in erg cm$^{-2}$ s$^{-1}$). {\it Bottom panel:} Optical--UV photometric observations. The large open circles represent MASTER unfiltered optical photometry reduced to the $V$ system while the points are $V$ ASAS-SN (nightly means) reduced to the {\it Swift} $V$ system. The filled circles show MASTER $V$-band photometry. The small open boxes correspond to the $U$ and $UVW1$ {\it Swift}/UVOT photometry.}
    \label{fig1}
\end{figure}

\begin{figure}
\includegraphics[scale=0.8,angle=0,trim=0 0 0 0]{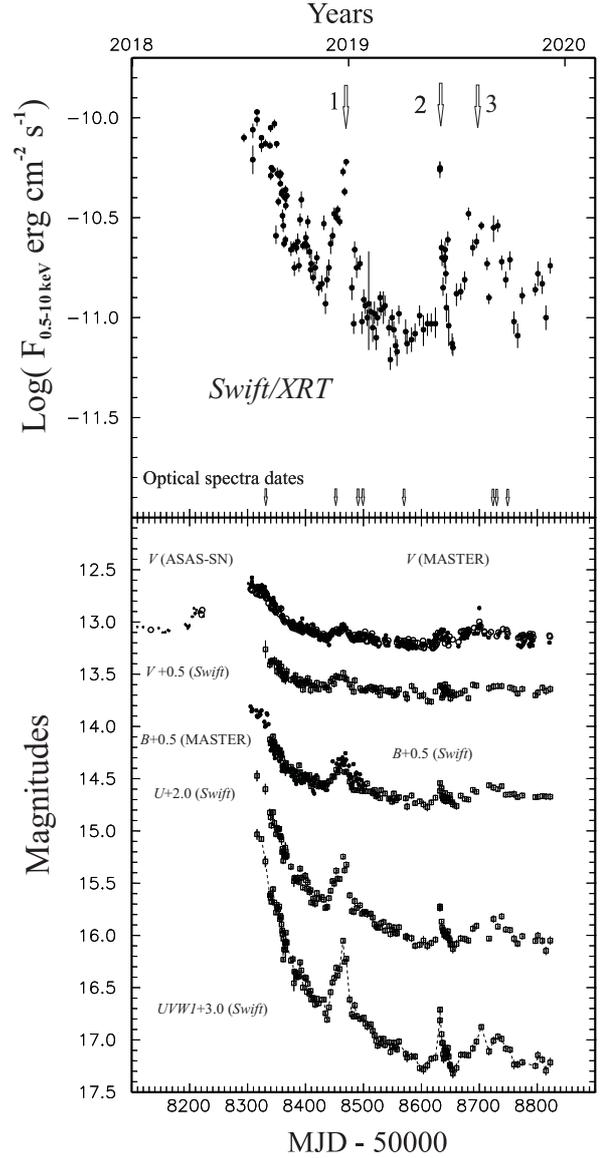}
 \caption{Multi-wavelength observations of NGC~1566 shown just for 2018-2019. {\it Top panel:} The {\it Swift}/XRT 0.5--10 keV X-ray flux (in erg cm$^{-2}$ s$^{-1}$). {\it Bottom panel:} The large open circles represent MASTER unfiltered optical photometry reduced to the $V$ system while the points are $V$-band ASAS-SN (nightly means) reduced to the {\it Swift} $V$ system. The filled circles show MASTER $BV$ photometry results. The open boxes correspond to the $UVW1$ and $UBV$ data obtained by {\it Swift}. The arrows indicate the Events 1, 2 and 3 (see text for the details).}
    \label{fig2}
\end{figure}

NGC~1566 has been a regular target of observation for the {\it Neil Gehrels Swift Observatory} \citep{Gehrels2004} for many years, starting in late 2007. Findings resulting from this monitoring programme have been published in a variety of studies \citep{Kawamuro2013, Ferrigno2018, Grupe2018, Grupe2018b, Oknyansky2019, Grupe2019, Parker2019}. In the present paper we have added to the analysis the most recent data (both from the XRT and UVOT telescopes), using the same methods as in Paper I and \cite{Oknyansky2017a}, but we uniformly re-reduced all available data to ensure usage of the most recent versions of the software and calibration files.  No significant variations were found between different versions of the reduced data. The new data include 51 dates for interval from 26th Aug.2019 till 5 Dec.2019. This allows us to trace the long and shot term  evolution of NGC~1566's behaviour including  post maximum period. 

The results obtained with the XRT and UVOT telescopes are shown in Fig.~\ref{fig1}--\ref{fig2} and will be discussed below in sections 3.1 and 3.2.

\subsection{Observations with the MASTER network}
Details on the MASTER \citep{Lipunov2010}, photometry methods, comparison stars, reductions and references can be found in Paper I. New observations (230 new dates) 
include interval after 25th Aug. 2018 till 5th Dec.2019.

The MASTER observation results are presented in Fig.~\ref{fig1} and \ref{fig2}. There we also show the ASAS-SN \citep[All-Sky Automated Survey for Supernovae, ][]{Shappee2014, Kochanek2017, Dai2018} $V$-band magnitudes reduced to the $V$ {\it Swift}/UVOT system. These light curves are discussed below in section 3.2.

\subsection{Optical spectral observations and reductions}

\begin{figure*}
\includegraphics[scale=1.0,angle=0,trim=0 0 0 0]{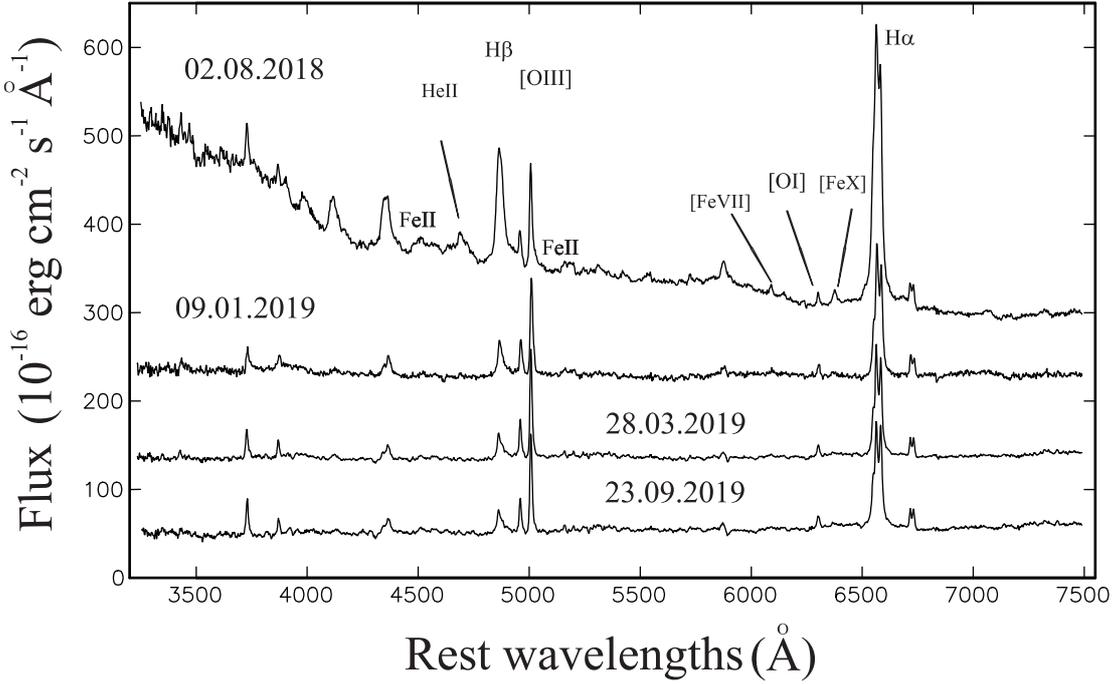}

    \caption{The isolated nuclear (low-resolution) non-stellar spectra of NGC~1566 (for 2 Aug. 2018, 9 Jan. 2019, 28 Mar. 2019 and 23 Sep. 2019) obtained by subtraction of the host galaxy spectrum  from the original spectrum. Some spectra are shifted up by $2\times F_c$ (28 Aug. 2018 and 9 Jan. 2019) and $F_c$ (28 Mar. 2019) respectively for display purposes (where $F_c = 10^{-14}$ erg~cm$^{-2}$ s$^{-1} \AA^{-1}$).}
    \label{fig3}
\end{figure*}

\begin{figure*}
	\includegraphics[scale=1.0,angle=0,trim=0 0 0 0]{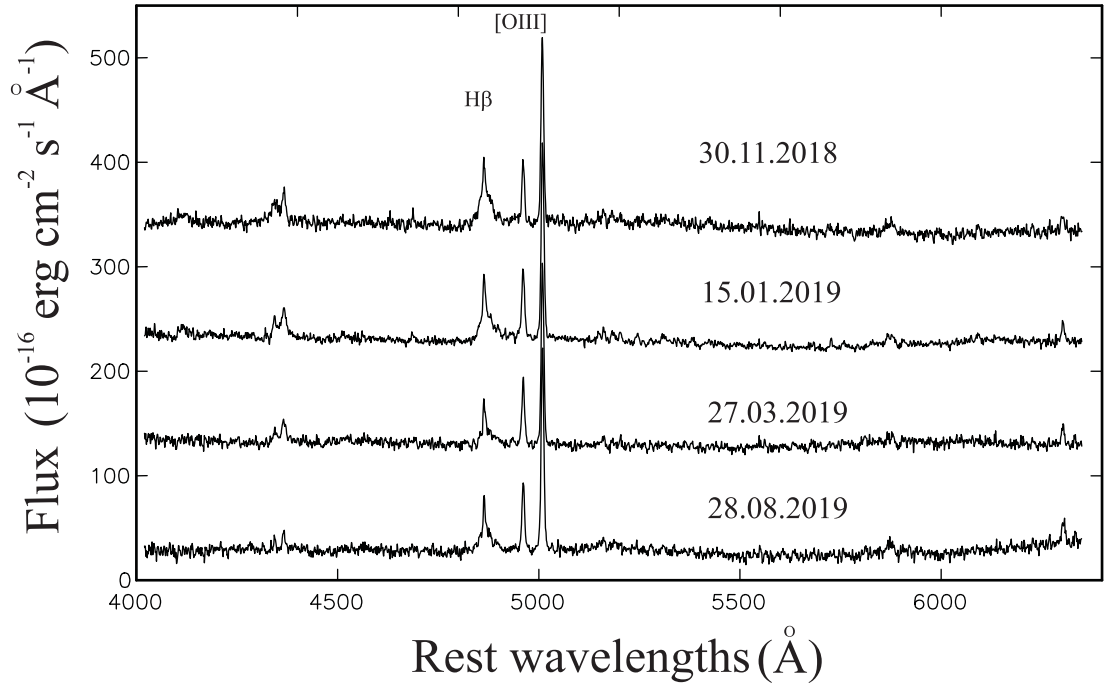}

    \caption{The isolated nuclear (medium-resolution) non-stellar spectra of NGC~1566 obtained by subtraction of the host galaxy spectrum from the original spectrum (see details in the text).
    Some spectra are shifted up by $3\times F_c$ (30 Nov. 2018), $2\times F_c$ (15 Jan. 2019) and $F_c$ (27 Mar. 2019) respectively for display purposes (where $F_c = 10^{-14}$ erg~cm$^{-2}$ s$^{-1} \AA^{-1}$).}
    \label{fig4}
\end{figure*}

Low- and medium-resolution spectra of NGC~1566 were obtained on 10 nights over the period Aug 2, 2018 to Sep 23, 2019 (see Table~\ref{tab1}) with the 1.9 m telescope at SAAO in Sutherland. The  August 2 spectrum was published in Paper I. 
We used the SpUpNIC Cassegrain spectrograph \citep{Crause2019} with low (300 grooves per mm) and medium (500 grooves per mm) resolution gratings.  The former span the range 3300 to 8500\AA~ with a nominal resolution of about 7 \AA, while the range for the latter was typically 3600 to 6400\AA~ with a nominal resolution of about 4 \AA. The spectrograph slit was oriented east-west. On each occasion at least two spectra were recorded and averaged into a single spectrum. Integration times and slit widths are given in Table~\ref{tab1}.
 All spectra from Aug 2, 2018 to Sep 23, 2019 were reprocessed and measured in a uniform way.
The low-resolution spectrum from Dec 1, 2018 could not be calibrated since the associated standard star observation could not be carried out.  We use this spectrum just for visual analysis of emission lines.

\begin{table}
\centering
\caption{Summary of optical spectral observations.}
\tabcolsep=0.11cm
\begin{tabular}{rcccc} \hline
 Date & Resolution & Integration & Slit & Flux   \\
     &        & time (s) & (\arcsec) & calibration \\ \hline \smallskip
 2 Aug. 2018 &  low & $2\times600$ & 2.7 & yes       \\ \smallskip
30 Nov. 2018 & medium & $2\times600$ & 0.9 &  yes  \\ \smallskip
 1 Dec. 2018 & low & $2\times600$ & 2.7 & no          \\ \smallskip
 1 Dec. 2018 & medium & $2\times600$ & 0.9 & yes    \\ \smallskip
 9 Jan. 2019 & low & $3\times600$ & 2.7 & yes            \\ \smallskip
15 Jan. 2019 & medium & $2\times1200$ & 0.9 & yes  \\ \smallskip
27 Mar.  2019 & medium & $2\times600$ & 0.9 & yes  \\ \smallskip
28 Mar.  2019 & low & $2\times600$ & 2.7 & yes         \\\smallskip
28 Aug. 2019 & medium &   $2\times1200$ & 0.9& yes \\\smallskip
 3 Sep. 2019 & low &  $2\times600$ & 2.7 & yes \\\smallskip
23 Sep. 2019 & low &  $2\times600$ & 2.7 & yes \\
\hline
\end{tabular}
\label{tab1}
\end{table}

Here we use the same methods for subtracting off-nuclear spectra as in Paper~I, where details can be found. The isolated nuclear non-stellar spectra in NGC~1566 were obtained by subtraction of the background galaxy starlight, estimated by suitably scaling an off-nuclear spectrum of the same galaxy, from the spectra displayed in Fig.~\ref{fig3} (low resolution) and in Fig.~\ref{fig4} (medium).  These spectra and the results will be discussed below in section 3.3.

\section{Results}

\subsection{{\it Swift}/XRT results}

The  lightcurve  in the 0.5--10 keV band (spanning the time period from Dec 11, 2007 to Dec 4, 2019) is shown in Fig.~\ref{fig1} (top panel) and for just 2018-2019 in Fig.~\ref{fig2} (top panel). The variability of the source observed in 2007--2018 was discussed in Paper~I.
Between 2007-2015 {\it Swift}/XRT observations were made only a few times per year. From these we can only conclude that the X-ray flux was decreasing to the minimum  in 2014--2015. After 2015 there was unfortunately a gap of about 3 years in the observations.  Intensive monitoring was started in 2018 just after the discovery of 
 the strong X-ray outburst in June 2018 (see details in Paper~I). After a maximum was reached at the beginning of July 2018, the fluxes in all bands decreased significantly, but  some re-brightenings were observed in MJD ranges 58440-58494 (hereafter referred to as ``Event 1") and 58603-58654 (``Event 2") \citep{Grupe2018b, Grupe2019}. One more re-brightening (``Event 3") was observed on Aug 5, 2019 mostly from the MASTER data, but on the nearest date of X-ray observations (Aug 8) NGC~1566 was relatively bright, both in X-rays and {\it UVW1}. Unfortunately, observations in other UVOT bands were not made at this date.

The minimum X-ray flux level recorded during 2018-2019 of about (6.2$\pm$0.1)$\times$ $10^{-12}$ ~erg~cm$^{-2}$ s$^{-1}$  was observed on Mar 3, 2019 (MJD 58546), while the most significant maximum flux measured on Jul. 16, 2018 (MJD 58316)  was (1.07$\pm$0.04$)\times$ $10^{-10}$ ~erg~cm$^{-2}$ s$^{-1}$ -- i.e., a decline of a factor of  17. That is significant variability, but smaller than the brightening by a factor of 50 times reported in our previous paper (Paper I) from a minimum in 2014 to Jul. 2018. As can be seen from Fig.~\ref{fig1} and Fig.~\ref{fig2}, after the minimum in Mar. 2019 the X-ray flux started increasing (if we remove from consideration the rapid brightenings in Event 2 and Event 3). The X-ray flux level observed on Sep 5, 2019 of 2.9$\pm0.2$ $\times$ $10^{-11}$~erg~cm$^{-2}$ s$^{-1}$ was about 4.7 times higher than it was on Mar 3, 2019. The spectral photon index remained $\sim$1.5 between Mar. and Sep. 2019. This indicates that the X-ray spectrum had become harder than it was at its maximum flux state in Jul. 2018 when the spectral photon index reached the value of about 2.2.

Our analysis in Paper I revealed a strong dependence of the spectral photon index on the source luminosity -- meaning that the amplitude of the X-ray variability was higher for soft X-rays.  The strong flux decline observed during the roughly 9 months after the maximum of Jul 2018 was about  30 times in the soft band (0.5--2.0 keV) and about  20 times in the hard band (2-10 keV). So we again see that the variations are higher in soft X-rays than in hard X-rays.  
Meanwhile, during the slow growth from Mar. to Sep.  2019 we didn't find any significant difference in the amplitudes of soft and hard X-ray fluxes.

We estimate Eddington ratios for the minimum (2014) and maximum (2018) using the 2--10 keV fluxes and a mass of $\sim10^{6.9}$ M$_{\sun}$ \citep{Woo2002}, a distance of $\sim$7.2 Mpc and taking a bolometric correction factor of 20 \citep{Vasudevan2009}. The obtained Eddington ratios are then $\sim$0.055\% for 2014 and $\sim$2.8\% for 2018.

\subsection{{\it Swift}/UVOT and MASTER photometry compared with {\it Swift}/XRT results}

Light curves in the optical {\it VBU} and UV bands are well correlated as can be clearly seen in Fig.~\ref{fig1} (bottom panel) for 2007-2019 and Fig.~\ref{fig2} (bottom panel) for 2018--2019.
The UV/Opt variation are mostly in agreement with X-ray variations for the minimum at 2014-2015 and the significant brightening to the maximum reached in July 2018. This brightening was started about 9 months before the maximum from the end of 2017 as seen in the ASAS-SN data \citep{Dai2018, Parker2019}. This was also independently confirmed by our MASTER photometry. The rise and decay times were about the same duration, viz. $\sim$9 months.
The variations between the maximum and minimum brightness during the previous year were consequently about $0.^{m}5$ in $V$, $1.0^{m}$ in $B$, $1.5^{m}$ in $U$ and $2.5^{m}$ in $UVW1$.

These differences for the UV/Opt bands can be explained at least in part by a more significant contribution from the host galaxy at longer wavelengths. We suspect that for the 2014 minimum most of the optical and UV fluxes in the aperture were from the host galaxy. So we can use the magnitudes at minimum light to estimate the fluxes from the host galaxy in the observed bands and to estimate partial amplitudes in each band for the nucleus during the drop from the maximum in Jul 2018 until Mar. 2019. We find that these amplitudes were similar, with variations of factors of $\sim$ 9 times in both UV and optical (UBV) bands.

As can be seen in Fig.~\ref{fig1} and Fig.~\ref{fig2}, all the variations in the optical and UV correlate well with the X-ray flux variations. After the maximum reached at the beginning of July 2018 the fluxes in all bands decreased quite rapidly, but with some fluctuations. The magnitude of the decline is largest for X-rays and decreases with increasing wavelength. There are several additional differences between  X-ray and Opt/UV variability which should be noted. Firstly, day-to-day X-ray flux fluctuations were observed, on occasions with amplitudes much larger than for UV and optical rapid variations on the same time scale. The difference  might be due to a much lower constant host galaxy contribution to X-rays and relatively smaller size of the X-ray emitting region. Some uncorrelated variations in X-ray and UV are present and might be connected in part with a possible  small time lag, smaller than the resolution of the observations. The lags between X-ray and UV/Optical variations  were  found for some tens of other AGNs \citep{Buisson2017} and usually these time delays are several times longer than what is predicted for a thin accretion disk by \cite{Shakura1973}. If we take into account these results and that the lags are $\tau\sim R/c\sim \dot{m}^{1/3}M^{2/3}$, then the expected delay values for NGC~1566 can be estimated to be less than 1 day. This estimate is in agreement with our analysis of the light curves, but the details will be discussed in a future paper.

\begin{figure}
\centering
\includegraphics[scale=0.55,angle=0, trim = 0 0 0 0]{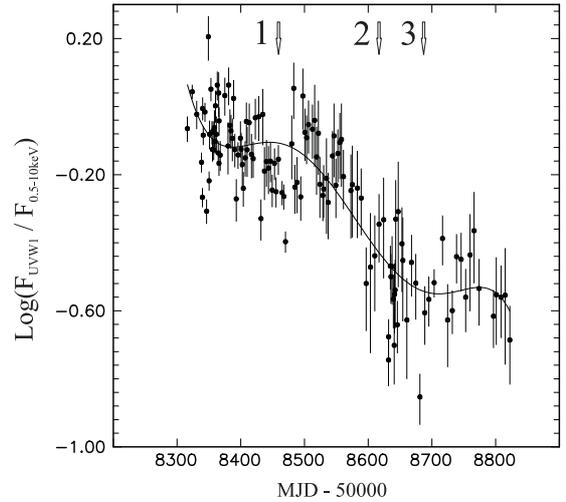}
 \caption {Variations of $F$(UVW1)/$F$(X-ray) during 2018-2019. The solid line is a fifth-order polynomial approximation. The arrows indicate the Events 1, 2 and 3 (see text for more details).}
    \label{fig5}
\end{figure}

The minimum in the optical and UV bands was reached not in  Mar. 2019, as for X-ray, but about 3 months later (if we do not take into account Event 2). From 3 Mar. -- 16 Jun. 2019 the flux in $UVW1$ decreased by a factor of $\sim$2 times, in $U$ by $\sim$1.7 and in $B$ by $\sim$1.2, while in $V$ it stayed about the same. This difference could be due to variations in obscuration which have the biggest effect in the UV, and have no effect for X-ray flux, which grew during this interval. The variations of $R=L_{\rm UV}/L_{\rm X-ray}$ are shown in Fig.~\ref{fig5}. The values $L_{\rm X-ray}$ and $L_{\rm UV}$ were calculated the same way as described by \cite{Ruan2019}. The host-galaxy contamination to the UVW1 was also removed the same way. As can be seen from Fig.~ \ref{fig5}, the ratio $R$ decreased significantly after Mar. 2019. A possible interpretation of this is given in the Discussion below.

\begin{figure}
\centering
\includegraphics[scale=0.65,angle=0, trim = 0 0 0 0]{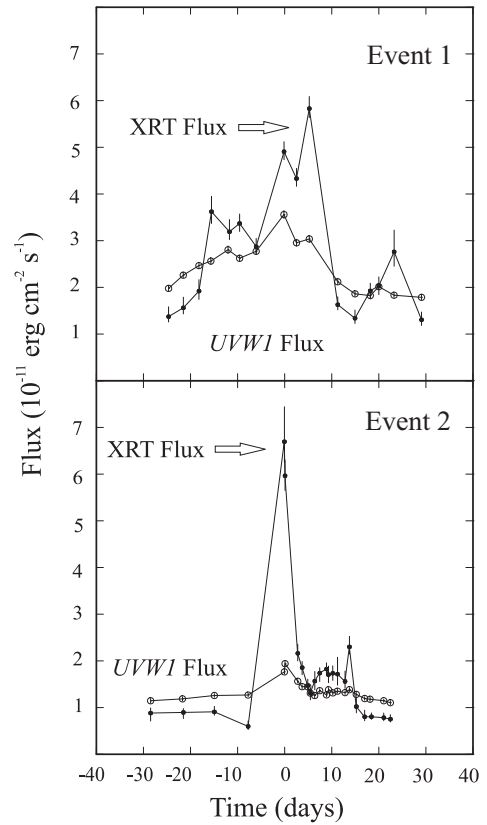}
 \caption {Light curves for $UVW1$ and XRT (in fluxes) during Event 1 and Event 2. Zero time in both cases corresponds to the maximum in the UV light curves.}
    \label{fig6}
\end{figure}

\begin{figure}
\centering
\includegraphics[scale=0.60,angle=0, trim = 0 0 0 0]{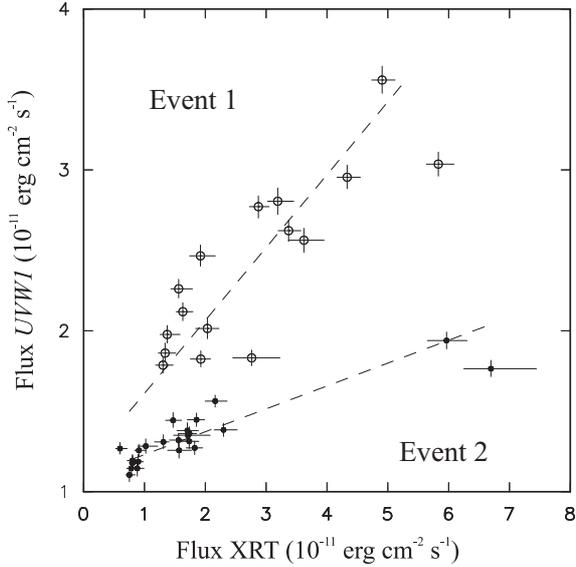}
 \caption {Comparison of the $UVW1$ flux to the XRT flux in Events 1 and 2.}
    \label{fig7}
\end{figure}

The most significant re-brightening phases in the light curves  (Events 1-3) seen in X-ray are also present in the UV and optical light curves. The  levels of the re-brightening reached in the UV and optical bands during Event 1  are significantly higher than for Event 2 (see Fig.~\ref{fig6} and Fig.~\ref{fig7}). This contrasts with the X-ray variations for which fluxes in the maxima were about the same.  The X-ray and UV variations for Events 1 and 2 correlate well, but the regression lines are very different for each event (see Fig.~\ref{fig7}) and that can reduce the correlation if the data are combined. The reason for the difference between Event 1 and Event 2 can be explained by a significantly shorter duration of X-ray outburst of Event 2 and a relatively larger size of the UV/Optical region compared to the X-ray one. Also, this difference is correlated with a decrease of the ratio $R$  between Events 1 and 2. We do not show the relation between the X-ray and UV fluxes for Event 3 since the maximum at Aug. 5 2019 was missed by {\it Swift}. This event was also very short, but the amplitude in $V$  was brighter than for Events 1 and 2.

\subsection{Optical spectrum: new results}

In Paper I we described in detail our spectrum obtained on Aug 2, 2018, which revealed a dramatic strengthening of the broad Balmer emission lines, the \ion{Fe}{ii} emission  lines, high-ionization coronal lines and the UV continuum compared to what had been published earlier. This result confirmed that NGC~1566 can be called a CL AGN.  In Paper I as well as here we assume that there are no significant variations of the NLR emission lines on a time scale of 1-2 years. This assumption is usual in most publications on the spectral variability of AGNs including some CL AGNs \citep[see e.g.,][]{Cohen1986,Shapovalova2019}. To date just a few extreme events of the NLR emission lines variability were reliably proven, and these were on longer time scales \citep[see e.g.,][]{Denney2014}. Our data for NGC 1566 in this paper shows significant correlation of the permitted line variations with those of the continuum, which makes it unlikely that the relative intensity variations we highlight in this paper are connected or significantly affected by possible variability of the forbidden lines. 

\begin{figure}
	\includegraphics[scale=0.57,angle=0,trim=0 0 0 0]{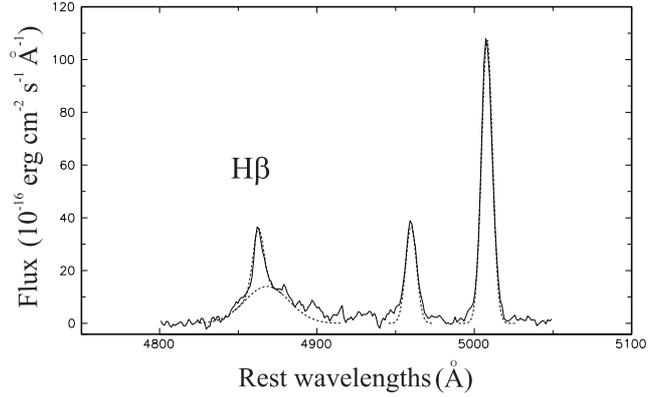}
    \caption{The nuclear  spectrum of 15 Jan 2019 with the estimated continuum subtracted (solid line) in the H$\upbeta$ and [\ion{O}{iii}] $\lambda$5007 region.  The narrow and broad line Gaussian component fits are shown by dashed lines.}
    \label{fig8}
\end{figure}

We have recalibrated all spectra (including data in Paper I) and remeasured  relative line intensities in a uniform way. We focus here on the spectral changes since the maximum spectrum of 2 Aug. 2018. Relative line intensities are presented in the Table~\ref{tab2}. The values  listed here (noted by $*$) include the contributions of narrow components as well as blended lines: [\ion{S}{ii}] $\lambda$6716,6731, [\ion{N}{ii}] $\lambda$6584,6584 for H$\upalpha$, [\ion{O}{iii}] $\lambda$4363 for H$\gamma$, possible \ion{Fe}{ii} contamination for \ion{He}{ii} $\lambda$4686 and [\ion{O}{i}] $\lambda$6363 with [\ion{Fe}{x}] $\lambda$6374. 

The emission from the HII region near the nucleus \citep{Silva2017} does not affect our analysis (see also the discussion on this point in Paper I). This can be seen when we fit multiple Gaussians to the lines (see Fig.~\ref{fig8}). The plot shows that just one Gaussian is sufficient to fit the narrow lines, and the red-shifted HII-region component evident in the spectra shown in the  \citet{Silva2017} paper is not visible here. We further note that the broad $H{\upbeta}$ component cannot be fitted well by a single Gaussian, which is not unexpected given the complex broad-line profiles associated with most AGNs.

\begin{figure}
	\includegraphics[scale=0.58,angle=0,trim=0 0 0 0]{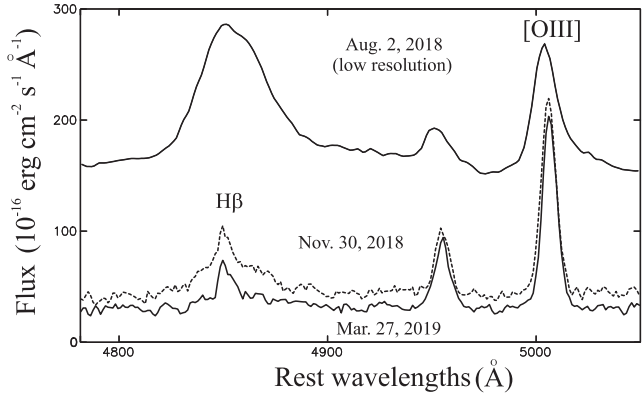}

    \caption{The isolated nuclear non-stellar spectra in NGC~1566 obtained by subtraction of the host galaxy spectrum from the original low-resolution spectrum from Aug 2, 2018  and the medium-resolution spectrum from Nov 30,  2018 (dashed line) and  Mar 27, 2019 in just H$\upbeta$ and [\ion{O}{iii}] $\lambda$5007 region.}
    \label{fig9}
\end{figure}

In view of the relatively low spectral resolution, and also because the broad lines are dominant, the narrow Balmer line strengths could not always be established to a reasonable accuracy. The contributions of the narrow components of the  Balmer lines were therefore generated by adopting the [\ion{S}{ii}]/H$\upalpha$(narrow)=0.78, [\ion{N}{ii}]$\lambda$6584/H$\upalpha$(narrow)=1.18  ratios  determined by \cite{Silva2017} and [\ion{N}{ii}]$\lambda$6584/[\ion{N}{ii}]$\lambda$6548=3.07 ratio from \cite{Storey2000}.  
To get the true [\ion{Fe}{x}]$\lambda$6374/[\ion{O}{i}]$\lambda$6300 ratios we subtract 0.3, which is the approximate theoretical value of [\ion{O}{i}] $\lambda$6363/[\ion{O}{i}]$\lambda$6300 \citep[see e.g.,][]{Storey2000}, from the measured values ([\ion{Fe}{x}] $\lambda$6374+[\ion{O}{i}] $\lambda$6363)/[\ion{O}{i}]$\lambda$6300.  As it is seen from Fig.~\ref{fig3},~\ref{fig4},~\ref{fig9} and Table~\ref{tab2}, our new spectra from 30 Nov. 2018 to 23 Sep. 2019 demonstrate a dramatic dimming of the broad Balmer lines, \ion{Fe}{ii}, the UV continuum and the coronal lines \ion{Fe}{vii} and \ion{Fe}{x} during the $\sim$4 months after the maximum. Subsequently the lines were very faint during Dec. 2018-2019, although some fluctuations were noted, which probably correlated with fluctuations of the UV and X-ray fluxes. For example, the broad Balmer lines and [\ion{Fe}{x}] line were a little bit stronger in Jan., Aug., Sep. 2019, reaching a minimum level in Mar. 2019, followed by partial re-brightenings (Event 1 and Event 3) and then a minimum level in Mar. 2019. Intensities of the H$\upbeta$, \ion{He}{ii} $\lambda$4686 and [\ion{Fe}{x}] $\lambda$6374 lines decreased by about a factor of five between Aug. 2018 and Mar. 2019. If we take into account that there is a narrow component of H$\upbeta$, which is not variable and has an intensity about 0.16 of [\ion{O}{iii}] $\lambda$5007 \citep{Silva2017}, then the amplitude of variability for the broad H$\upbeta$ component has to be about a  factor of 6-8. Some part of [\ion{Fe}{x}] $\lambda$6374 is also radiated from a large distance from the central source \citep[see e.g.,][]{Oliver2005}.  This means that the part of the coronal line which is radiated from a smaller distance has  an amplitude of variability of more than a factor of five. The H$\upalpha$/H$\upbeta$ ratio changed from $\sim$2.7 in high state (2 Aug. 2018) to $\sim$6.0 in the low state (28 Mar. 2019). If we take into account that this calculated value is very sensitive to the choice of  narrow $H{\upbeta}$ component, as well as to some other possible systematic errors, then the uncertainty might reach $\sim30\%$ when the AGN is in a low state. These results will be discussed below.

\begin{table*}
\centering
\caption{Measured emission line intensity ratios ($*$ These values include the contributions of narrow components as well as blended lines. See text for the details).}
\tabcolsep=0.11cm
\begin{tabular}{ccccccccccc} \hline
Line &&&& Date &&&&\\
(includes blends) &02.08.18&30.11.18&01.12.2018&09.01.19&15.01.19&27.03.19&28.03.19&28.08.19&04.09.19&23.09.19 \\ \hline \smallskip
 $*$ H$\upbeta$/[OIII]5007    &  2.92 & 1.01  & 0.98 & 0.77 & 0.83 & 0.55 & 0.51& 0.72 &0.64 &0.66 \\ \smallskip
 $*$ [SII]/[OIII]5007  &  0.38 & -         & -       & 0.23 & -       & -        & 0.30&-&0.34&0.27\\ \smallskip
 $*$ H$\upalpha$/H$\upbeta$   &  2.8 & -         & -       & 3.9 & -       & -        & 5.7&-&5.8&6.1\\ \smallskip
 H$\upalpha$/H$\upbeta$   &  2.6 & -         & -       & 3.7 & -       & -        & 5.5&-&5.4&6.2\\ \smallskip
  $*$ H$\gamma$/H$\upbeta$ &  0.52 & 0.44 & 0.57 & 0.61 & 0.50 & 0.53 & 0.48&-&0.41&0.46 \\ \smallskip
 $*$ HeII4685/H$\upbeta$      & 0.28 & -   & - & -       & - & - & -  &-&-&-    \\ \smallskip
 $*$ [FeX]/[OI]6300     &  2.2  & -          & -       & 0.8  & 1.0     & 0.6  & 0.4&-&0.6&0.9 \\ \smallskip
 [FeX]/[OI]6300     &  1.9  & -          & -       & 0.5  & 0.7     & 0.3  & 0.1&-&0.3&0.6 \\ 
\hline

\end{tabular}

\label{tab2}
\end{table*}

\section{Discussion}

\subsection{The multi-wavelength variability}
Overall, the X-ray, UV/Opt variability of NGC 1566 is not unusual for a CL AGN. The dramatic changes of factors 1.5 orders of magnitude in the X-ray flux along with correlated variations at UV/Opt wavelengths is a typical property of CL AGNs. However, whilst a strong correlation of the X-ray and UV/Opt is not common among  AGNs \citep[see e.g.,][] {Edelson2000, Buisson2017}, there are exceptions for some other CL AGNs, viz. NGC~ 2617, NGC~4151 and NGC~5548 \citep[see e.g.,][]{Shappee2014, Oknyansky2017a, Edelson2017, Mchardy2014}. From these results we can conclude that the variability across several wavebands in CL AGNs (spanning from X-rays to the  UV/Optical) is driven by variable illumination of the accretion disc (AD) by soft X-rays. We have to determine the most significant common features of such variability.  This can help us to find the most likely explanations for the CL phenomenon (see e.g. discussion and references in Paper I and below).  There are several different possibilities that have been considered: variable obscuration, disk flares, tidal disruption events (TDEs), and supernova events.

The duration  of the rise and decay  times  for the  main brightening in 2018 was about the same ($\sim 9$ months) and much longer than the recurrent brightenings (with lower amplitude) described by \cite{Alloin1986} ($\sim$20 days).  A similar dramatic brightening was observed previously only in 1962, with re-brightening about one year later \cite{Quintana1975}. Such re-brightenings soon after the main maximum may be common in the variability of CL AGNs.   

Whilst there is good evidence for obscuration in a very high percentage of AGNs (see \citealt{Gaskell17} and references therein), it is not the dominant variability mechanism.  It is difficult to explain the dramatic rise of the X-ray and UV/optical flux by a reduction in obscuration. The main problem with the obscuration theory is that it would predict much longer time scales for the spectral changes than what is observed in CL AGNs \citep[see e.g.,][]{Sheng2017}. For NGC 1566 we expect timescales of more than a few years for a CL event in the obscuration scenario, which is much longer than what is observed.  It does not exclude the possibility of intervening clouds causing variable obscuration, but seems to exclude that as a dominant explanation for the CL mechanism. The more obvious explanation of this, and AGN variability in general, is that it is due to variation of the intrinsic energy-generation rate (for example, violent flares above the accretion disc).  Since such variability is seen in all AGNs, CL AGNs are just extreme examples of this.  The timescales of variability of CL-AGNs at various wavelengths are similar to the variability timescales of AGNs in general.  

These two leading possibilities -- intrinsic variability and variable obscuration -- do not need to be mutually exclusive. For example, a dramatic change in the energy generation could cause the sublimation of dust in clouds near the central source energy source \citep{Oknyansky2017a, Oknyansky2019a}. If such sublimation occurs in some clouds along the line of sight, then the rise in the UV can be explained in part with the change in obscuration. The typical time for the recovery of dust clouds after the UV flux has abated can be several years \citep{Oknyansky2017a, Oknyansky2019a,  Kishimoto2013}, but it could be less for NGC~1566 if we take into account the significantly smaller mass of the SMBH (super massive black hole) compared with NGC~4151. So we could see some increase in obscuration during the time when the energy generation is falling.  

We do see some differences in variability in different wavelengths after Mar. 2019, which can be explained by variable obscuration since it must be most significant for the UV and less in optical bands, and with hardly any noticeable effect for X-ray flux.  There are other possible explanations for this difference in X-ray and UV variability. For example, the strong change in the UV to X-ray flux  ratio observed after Mar. 2019 can be connected with changing of the height of the X-ray source that can reduce the amount of UV/optical emission produced by reprocessing in the accretion disk \citep[see e.g.,][]{Breedt2009}. Also a transition in accretion mode might explain the observed differences in properties of  variability in  high and low states \citep[see e.g.,][]{Liu2020}

The most intriguing question is what the mechanism would be for such dramatic switching on or off of the energy generation of CL AGNs. This has been explored in some firsts reports on CL events as well as in recent papers \citep{Lyuty1984, Penston1984, Runnoe2016, Katebi2019}. The most popular idea relates to disc instabilities. One of such mechanism of instability \citep{Lightman1974} was discussed by \cite{Parker2019} in connection with NGC~1566: that the inner disk is in a low state until radiation pressure exceeds gas pressure. If that happens then an accretion disk switches on to a high state. This mechanism is consistent with the time scales of variability, but cannot explain our result that the rise time is not longer than the decay one. The radiation pressure instability in relation to NGC~1566 were also considered by \cite{Sniegowska2019} to explain recurrent outbursts observed in 1970-1990. This type of variability has not, however, been seen during the last 12 years. The most significant outbursts recur in NGC~1566 on a time scale several decades (in 1962 and 2018). So we need some mechanism which can explain why the object is in a low state most of the time when the broad component of H$\upbeta$ is almost undetectable but then on rare occasions we get dramatic brightenings. The idea that accretion rate transitions similar to those observed in black hole binaries may  work for SMBH has previously been mooted \citep[see e.g.,][]{Liu2020}. However, the then expected time scale of several years is longer than rise and decay times observed in NGC~1566.
The TDE and supernova explanations can be rejected given the differences between expected and observed time scales for such events, as well as their spectral properties and evolution. An alternative theory proposes the tidal stripping of stars \citep{Campana15, Ivanov-Chernyakova06}, which could lead to more frequent and recurrent events \citep{Komossa17} if these stars have bound orbits similar to some known objects near the SMBH in the Milky Way. Also, these stellar striping events could trigger accretion disk instabilities and explain short time-scales for brightening, as well as why re-brightenings would recur \citep[see e.g.,][]{Ivanov2018}. However, the mechanism has not been investigated in sufficient detail yet.  More references on the possible mechanisms of the CL events can be found in discussions by \cite{MacLeod2019,Runnoe2016, Ruan2019}.  At present we are far from understanding the CL phenomenon, and many questions remain. 

\subsection{The optical spectral variability}

The strong variability of UV flux in NGC~1566 should affect the ionisation conditions in the emission regions near the central source and explain the strong variability of the broad emission lines. Variability of the broad Balmer and \ion{He}{ii} $\lambda$4686 lines are typical for CL AGNs. The intensities of these lines are correlated with the level of the UV/optical continuum.  
Possible time delays in the lines variability depend from the size of BLR. \cite{Alloin1985} and \cite{Silva2017} estimated the size, between 1 and 15 light-days and less than 20 light-days, respectively. We do not have enough spectral epochs to perform a reverberation mapping measurement of the possible lag in H$\upbeta$ variability relative to UV, but we estimate the delay as 2 days (or less) from the radius -- luminosity relationship determined from other AGN \citep[see e.g.,][]{Shappee2014}. Consistent with this model, we found that the amplitude of this variability decreases in NGC~1566 as one goes from \ion{He}{ii}$\lambda$4686 to H$\upbeta$ and H$\upalpha$ (see Table~\ref{tab2}). (However we have take into account that lower amplitude of variability of H$\upalpha$ might in part due to blending with narrow \ion{N}{ii} lines.) Steep Balmer decrements such as the high H$\upalpha$/H$\upbeta$ ratio in the minimum (28 Mar. 2019) have been widely attributed to a combination of low optical depth and low ionisation parameter in the the BLR (broad line region) as  in case of Mrk~609, Mrk~883 and UGC~7064 \citep{Goodrich1995, Rudy1988}. The variability of the Balmer decrements is a typical property of a CL AGN and usually it is anti-correlated with the continuum flux variations \citep{Shapovalova2004}. For example, in the case of NGC~4151, it was measured to vary from 2 to 8 \citep{Shapovalova2008}. Variable obscuration is often considered for the explanation \citep[see e.g,][]{Gaskell17}, but there are other possibilities \citep[see e.g.,][]{Ilich2012}. 

One of the most interesting result found from our spectral data is the strengthening of the UV Balmer continuum  during the maximum of 2018 (Paper I) and its weakness during low states. This result can be seen independently from spectra obtained by SALT in 2012 and 2018 \citep{Kollatschny2020}. A similar effect has been detected in other CL AGNs as they change their states \citep[see e.g., ][]{Shappee2014, Edelson2000}, and it is probably a common property of CL AGNs.

The dramatic variability (approaching a factor of a hundred) of the soft X-ray flux in NGC~1566 has a strong effect on the ionization of gas near the central source and can thus explain the significant variability of high-ionisation lines like  [\ion{Fe}{vii}] and [\ion{Fe}{x}] which need strong soft X-ray fluxes for such high ionisation stages. The greater variability of these coronal lines in AGN compared to the majority of other forbidden lines could be due to the transition probabilities of coronal lines being several orders of magnitude higher than for lower-ionization forbidden lines \citep{Penston1984a}. 
This would also not be surprising since high-ionization coronal lines would be expected to arise in a smaller  region than other forbidden lines \citep[see e.g., Paper I,][]{Rose2015,Netzer1974, Osterbrock1982,Oknyansky1982, Chuvaev1989, Oknyansky1991, Veilleux1988, Landt2015, Landt2015b, Parker2016}.
In case of the NGC~1566 this variability is particularly obvious (see Paper I). It might even be a typical features of CL AGNs in general. However, the interpretation of the fast variability of coronal lines presents certain difficulties. The most promising hypothesis may be an assumption about the emission of these lines in polar cones near the accretion disk \citep{Oknyansky1988, Oknyansky1989}. If the accretion disk of NGC~1566 has an approximately face-on orientation \citep{Grupe2019, Combes2019} the coronal regions might be extended along the line of sight. In this case the region where the coronal lines are radiated can have a large enough volume to explain observed flux in the lines and at the same time the variability can be on very small time scales and without significant lags as has been proposed for dust reprocessing by \citet{Oknyansky2015}.

\section{Conclusions}

We have shown, from new  optical spectroscopy (1.9 m SAAO) and multi-wavelength photometry (MASTER, {\it Swift} Ultraviolet/Optical and  XRT Telescopes) of the AGN in NGC~1566, that after maximum was reached (at the beginning of July 2018) the fluxes in all bands dramatically declined with some fluctuations. Re-brightening events during the decline from maximum were observed in Dec. 2018 (Event 1), May 2019 (Event 2) and Aug. 2019 (Event 3). The amplitudes of X-ray fluxes in Event 1 and Event 2 were about the same, but UV/optical flux variations were significantly lower in the last case. The amplitude of the flux variability is strongest in the X-ray band and decreases in the UV and optical bands. We have found a strong decrease of the UV/X-ray ratio after Mar. 2019 and a rise of the|X-ray flux. The strength of the broad permitted, high ionisation [\ion{Fe}{x}] $\lambda$6374 lines and UV continuum dropped significantly up to the end of March 2019 and NGC~1566 can , since then, again be classified as a Seyfert 1.8 to 1.9. We have therefore established that NGC~1566 is a clear case of a CL AGN.  Our most recent spectra of Aug.-Sep. 2019 show lines again slightly stronger than in Mar. 2019.
We suspect that the most probable scenarios of such dramatic variability are connected with AD instabilities which also can be triggered by events like stellar striping. The last option can explain recurrent CL events.

\section*{Acknowledgements}
HW and FVW thank the South African Astronomical Observatory for the generous allocation of telescope time which also resulted in the spectra presented in this paper. We also express our thanks to the {\it Swift} ToO team for organizing and executing the observations. This work was supported in part by the Russian Foundation for Basic Research through grant 17-52-80139 BRICS-a and by the BRICS Multilateral Joint Science and Technology Research Collaboration grant 110480. MASTER work was supported by Lomonosov Moscow State University Development Programme and  the Russian Foundation for Basic Research through grant 17-52-80133 BRICS . DB is supported by the National Research Foundation of South Africa. We are grateful to H.~Netzer, K.~Malanchev, C.M.~Gaskell, E.~Bon, D.~Tsvetkov, J.M.~Wang  and P.~Ivanov   for useful discussions.



\bibliographystyle{mnras}
\expandafter\ifx\csname natexlab\endcsname\relax\def\natexlab#1{#1}\fi

\interlinepenalty=10000

\bibliography{1566} 

\begin{thebibliography}{}
\makeatletter
\relax
\def\mn@urlcharsother{\let\do\@makeother \do\$\do\&\do\#\do\^\do\_\do\%\do\~}
\def\mn@doi{\begingroup\mn@urlcharsother \@ifnextchar [ {\mn@doi@}
  {\mn@doi@[]}}
\def\mn@doi@[#1]#2{\def\@tempa{#1}\ifx\@tempa\@empty \href
  {http://dx.doi.org/#2} {doi:#2}\else \href {http://dx.doi.org/#2} {#1}\fi
  \endgroup}
\def\mn@eprint#1#2{\mn@eprint@#1:#2::\@nil}
\def\mn@eprint@arXiv#1{\href {http://arxiv.org/abs/#1} {{\tt arXiv:#1}}}
\def\mn@eprint@dblp#1{\href {http://dblp.uni-trier.de/rec/bibtex/#1.xml}
  {dblp:#1}}
\def\mn@eprint@#1:#2:#3:#4\@nil{\def\@tempa {#1}\def\@tempb {#2}\def\@tempc
  {#3}\ifx \@tempc \@empty \let \@tempc \@tempb \let \@tempb \@tempa \fi \ifx
  \@tempb \@empty \def\@tempb {arXiv}\fi \@ifundefined
  {mn@eprint@\@tempb}{\@tempb:\@tempc}{\expandafter \expandafter \csname
  mn@eprint@\@tempb\endcsname \expandafter{\@tempc}}}

\bibitem[\protect\citeauthoryear{{Alloin}, {Pelat}, {Phillips}  \&
  {Whittle}}{{Alloin} et~al.}{1985}]{Alloin1985}
{Alloin} D.,  {Pelat} D.,  {Phillips} M.,   {Whittle} M.,  1985, \mn@doi [\apj]
  {10.1086/162783}, \href {http://adsabs.harvard.edu/abs/1985ApJ...288..205A}
  {288, 205}

\bibitem[\protect\citeauthoryear{{Alloin}, {Pelat}, {Phillips}, {Fosbury}  \&
  {Freeman}}{{Alloin} et~al.}{1986}]{Alloin1986}
{Alloin} D.,  {Pelat} D.,  {Phillips} M.~M.,  {Fosbury} R.~A.~E.,   {Freeman}
  K.,  1986, \mn@doi [\apj] {10.1086/164475}, \href
  {http://adsabs.harvard.edu/abs/1986ApJ...308...23A} {308, 23}

\bibitem[\protect\citeauthoryear{{Baribaud}, {Alloin}, {Glass}  \&
  {Pelat}}{{Baribaud} et~al.}{1992}]{Baribaud1992}
{Baribaud} T.,  {Alloin} D.,  {Glass} I.,   {Pelat} D.,  1992, \aap, \href
  {http://adsabs.harvard.edu/abs/1992A%26A...256..375B} {256, 375}

\bibitem[\protect\citeauthoryear{{Breedt} et~al.,}{{Breedt}
  et~al.}{2009}]{Breedt2009}
{Breedt} E.,  et~al., 2009, \mn@doi [\mnras]
  {10.1111/j.1365-2966.2008.14302.x}, \href
  {https://ui.adsabs.harvard.edu/abs/2009MNRAS.394..427B} {394, 427}

\bibitem[\protect\citeauthoryear{{Buisson}, {Lohfink}, {Alston}  \&
  {Fabian}}{{Buisson} et~al.}{2017}]{Buisson2017}
{Buisson} D.~J.~K.,  {Lohfink} A.~M.,  {Alston} W.~N.,   {Fabian} A.~C.,  2017,
  \mn@doi [\mnras] {10.1093/mnras/stw2486}, \href
  {https://ui.adsabs.harvard.edu/abs/2017MNRAS.464.3194B} {464, 3194}

\bibitem[\protect\citeauthoryear{{Campana}, {Mainetti}, {Colpi}, {Lodato},
  {D'Avanzo}, {Evans}  \& {Moretti}}{{Campana} et~al.}{2015}]{Campana15}
{Campana} S.,  {Mainetti} D.,  {Colpi} M.,  {Lodato} G.,  {D'Avanzo} P.,
  {Evans} P.~A.,   {Moretti} A.,  2015, \mn@doi [\aap]
  {10.1051/0004-6361/201525965}, \href
  {http://adsabs.harvard.edu/abs/2015A%26A...581A..17C} {581, A17}

\bibitem[\protect\citeauthoryear{{Chuvaev} \& {Oknyanskii}}{{Chuvaev} \&
  {Oknyanskii}}{1989}]{Chuvaev1989}
{Chuvaev} K.~K.,  {Oknyanskii} V.~L.,  1989, \sovast, \href
  {https://ui.adsabs.harvard.edu/abs/1989SvA....33....1C} {33, 1}

\bibitem[\protect\citeauthoryear{{Cohen}, {Rudy}, {Puetter}, {Ake}  \&
  {Foltz}}{{Cohen} et~al.}{1986}]{Cohen1986}
{Cohen} R.~D.,  {Rudy} R.~J.,  {Puetter} R.~C.,  {Ake} T.~B.,   {Foltz} C.~B.,
  1986, \mn@doi [\apj] {10.1086/164758}, \href
  {https://ui.adsabs.harvard.edu/abs/1986ApJ...311..135C} {311, 135}

\bibitem[\protect\citeauthoryear{{Combes} et~al.,}{{Combes}
  et~al.}{2019}]{Combes2019}
{Combes} F.,  et~al., 2019, \mn@doi [\aap] {10.1051/0004-6361/201834560}, \href
  {https://ui.adsabs.harvard.edu/abs/2019A&A...623A..79C} {623, A79}

\bibitem[\protect\citeauthoryear{{Crause} et~al.,}{{Crause}
  et~al.}{2019}]{Crause2019}
{Crause} L.~A.,  et~al., 2019, \mn@doi [JATIS] {10.1117/1.JATIS.5.2.024007},
  \href {https://ui.adsabs.harvard.edu/abs/2019JATIS...5b4007C} {5, 024007}

\bibitem[\protect\citeauthoryear{{Dai}, {Stanek}, {Kochanek}, {Shappee}  \&
  {ASAS-SN Collaboration}}{{Dai} et~al.}{2018}]{Dai2018}
{Dai} X.,  {Stanek} K.~Z.,  {Kochanek} C.~S.,  {Shappee} B.~J.,   {ASAS-SN
  Collaboration} 2018, The Astronomer's Telegram, \href
  {http://adsabs.harvard.edu/abs/2018ATel11893....1D} {11893}

\bibitem[\protect\citeauthoryear{{Denney} et~al.,}{{Denney}
  et~al.}{2014}]{Denney2014}
{Denney} K.~D.,  et~al., 2014, \mn@doi [\apj] {10.1088/0004-637X/796/2/134},
  \href {https://ui.adsabs.harvard.edu/abs/2014ApJ...796..134D} {796, 134}

\bibitem[\protect\citeauthoryear{{Edelson}, {Gedney}, {Abbot}, {Dyer}  \&
  {Rolt}}{{Edelson} et~al.}{2000}]{Edelson2000}
{Edelson} G.~S.,  {Gedney} C.~J.,  {Abbot} P.~A.,  {Dyer} I.,   {Rolt} K.~D.,
  2000, \mn@doi [Acoustical Society of America Journal] {10.1121/1.428745},
  \href {https://ui.adsabs.harvard.edu/abs/2000ASAJ..107.2890E} {107, 2890}

\bibitem[\protect\citeauthoryear{{Edelson} et~al.,}{{Edelson}
  et~al.}{2017}]{Edelson2017}
{Edelson} R.,  et~al., 2017, \mn@doi [\apj] {10.3847/1538-4357/aa6890}, \href
  {https://ui.adsabs.harvard.edu/abs/2017ApJ...840...41E} {840, 41}

\bibitem[\protect\citeauthoryear{{Elagali} et~al.,}{{Elagali}
  et~al.}{2019}]{Elagali2019}
{Elagali} A.,  et~al., 2019, \mn@doi [\mnras] {10.1093/mnras/stz1448}, \href
  {https://ui.adsabs.harvard.edu/abs/2019MNRAS.487.2797E} {487, 2797}

\bibitem[\protect\citeauthoryear{{Ferrigno}, {Siegert}, {Sanchez-Fernandez},
  {Kuulkers}, {Ducci}, {Savchenko}  \& {Bozzo}}{{Ferrigno}
  et~al.}{2018}]{Ferrigno2018}
{Ferrigno} C.,  {Siegert} T.,  {Sanchez-Fernandez} C.,  {Kuulkers} E.,  {Ducci}
  L.,  {Savchenko} V.,   {Bozzo} E.,  2018, The Astronomer's Telegram, \href
  {http://adsabs.harvard.edu/abs/2018ATel11783....1F} {11783}

\bibitem[\protect\citeauthoryear{{Fitch}, {Pacholczyk}  \& {Weymann}}{{Fitch}
  et~al.}{1967}]{Fitch1967}
{Fitch} W.~S.,  {Pacholczyk} A.~G.,   {Weymann} R.~J.,  1967, \mn@doi [\apjl]
  {10.1086/180095}, \href
  {https://ui.adsabs.harvard.edu/abs/1967ApJ...150L..67F} {150, L67}

\bibitem[\protect\citeauthoryear{{Gaskell}}{{Gaskell}}{2017}]{Gaskell17}
{Gaskell} C.~M.,  2017, \mn@doi [\mnras] {10.1093/mnras/stx094}, \href
  {https://ui.adsabs.harvard.edu/abs/2017MNRAS.467..226G} {467, 226}

\bibitem[\protect\citeauthoryear{{Gehrels} et~al.,}{{Gehrels}
  et~al.}{2004}]{Gehrels2004}
{Gehrels} N.,  et~al., 2004, \mn@doi [\apj] {10.1086/422091}, \href
  {http://adsabs.harvard.edu/abs/2004ApJ...611.1005G} {611, 1005}

\bibitem[\protect\citeauthoryear{{Glass}}{{Glass}}{2004}]{Glass2004}
{Glass} I.~S.,  2004, \mn@doi [\mnras] {10.1111/j.1365-2966.2004.07712.x},
  \href {https://ui.adsabs.harvard.edu/abs/2004MNRAS.350.1049G} {350, 1049}

\bibitem[\protect\citeauthoryear{{Goodrich}}{{Goodrich}}{1995}]{Goodrich1995}
{Goodrich} R.~W.,  1995, \mn@doi [\apj] {10.1086/175256}, \href
  {https://ui.adsabs.harvard.edu/abs/1995ApJ...440..141G} {440, 141}

\bibitem[\protect\citeauthoryear{{Grupe}, {Komossa}  \& {Schartel}}{{Grupe}
  et~al.}{2018a}]{Grupe2018}
{Grupe} D.,  {Komossa} S.,   {Schartel} N.,  2018a, The Astronomer's Telegram,
  \href {http://adsabs.harvard.edu/abs/2018ATel11903....1G} {11903}

\bibitem[\protect\citeauthoryear{{Grupe} et~al.,}{{Grupe}
  et~al.}{2018b}]{Grupe2018b}
{Grupe} D.,  et~al., 2018b, The Astronomer's Telegram, \href
  {http://adsabs.harvard.edu/abs/2018ATel12314....1G} {12314}

\bibitem[\protect\citeauthoryear{{Grupe} et~al.,}{{Grupe}
  et~al.}{2019}]{Grupe2019}
{Grupe} D.,  et~al., 2019, The Astronomer's Telegram, \href
  {https://ui.adsabs.harvard.edu/abs/2019ATel12826....1G} {12826, 1}

\bibitem[\protect\citeauthoryear{{Ili{\'c}}, {Popovi{\'c}}, {La Mura}, {Ciroi}
  \& {Rafanelli}}{{Ili{\'c}} et~al.}{2012}]{Ilich2012}
{Ili{\'c}} D.,  {Popovi{\'c}} L.~{\v{C}}.,  {La Mura} G.,  {Ciroi} S.,
  {Rafanelli} P.,  2012, \mn@doi [\aap] {10.1051/0004-6361/201219299}, \href
  {https://ui.adsabs.harvard.edu/abs/2012A&A...543A.142I} {543, A142}

\bibitem[\protect\citeauthoryear{{Ivanov} \& {Chernyakova}}{{Ivanov} \&
  {Chernyakova}}{2006}]{Ivanov-Chernyakova06}
{Ivanov} P.~B.,  {Chernyakova} M.~A.,  2006, \mn@doi [\aap]
  {10.1051/0004-6361:20053409}, \href
  {http://adsabs.harvard.edu/abs/2006A%26A...448..843I} {448, 843}

\bibitem[\protect\citeauthoryear{{Ivanov}, {Zhuravlev}  \&
  {Papaloizou}}{{Ivanov} et~al.}{2018}]{Ivanov2018}
{Ivanov} P.~B.,  {Zhuravlev} V.~V.,   {Papaloizou} J.~C.~B.,  2018, \mn@doi
  [\mnras] {10.1093/mnras/sty2493}, \href
  {https://ui.adsabs.harvard.edu/abs/2018MNRAS.481.3470I} {481, 3470}

\bibitem[\protect\citeauthoryear{{Katebi} et~al.,}{{Katebi}
  et~al.}{2018}]{Katebi2018}
{Katebi} R.,  et~al., 2018, arXiv e-prints, \href
  {https://ui.adsabs.harvard.edu/abs/2018arXiv181103694K} {p. arXiv:1811.03694}

\bibitem[\protect\citeauthoryear{{Katebi} et~al.,}{{Katebi}
  et~al.}{2019}]{Katebi2019}
{Katebi} R.,  et~al., 2019, \mn@doi [\mnras] {10.1093/mnras/stz1552}, \href
  {https://ui.adsabs.harvard.edu/abs/2019MNRAS.487.4057K} {487, 4057}

\bibitem[\protect\citeauthoryear{{Kawamuro}, {Ueda}, {Tazaki}  \&
  {Terashima}}{{Kawamuro} et~al.}{2013}]{Kawamuro2013}
{Kawamuro} T.,  {Ueda} Y.,  {Tazaki} F.,   {Terashima} Y.,  2013, \mn@doi
  [\apj] {10.1088/0004-637X/770/2/157}, \href
  {http://adsabs.harvard.edu/abs/2013ApJ...770..157K} {770, 157}

\bibitem[\protect\citeauthoryear{{Kishimoto} et~al.,}{{Kishimoto}
  et~al.}{2013}]{Kishimoto2013}
{Kishimoto} M.,  et~al., 2013, \mn@doi [\apjl] {10.1088/2041-8205/775/2/L36},
  \href {https://ui.adsabs.harvard.edu/abs/2013ApJ...775L..36K} {775, L36}

\bibitem[\protect\citeauthoryear{{Kochanek} et~al.,}{{Kochanek}
  et~al.}{2017}]{Kochanek2017}
{Kochanek} C.~S.,  et~al., 2017, \mn@doi [\pasp] {10.1088/1538-3873/aa80d9},
  \href {http://adsabs.harvard.edu/abs/2017PASP..129j4502K} {129, 104502}

\bibitem[\protect\citeauthoryear{{Komossa} et~al.,}{{Komossa}
  et~al.}{2017}]{Komossa17}
{Komossa} S.,  et~al., 2017, in {Gomboc} A.,  ed.,  IAU Symposium Vol. 324, New
  Frontiers in Black Hole Astrophysics. pp 168--171

\bibitem[\protect\citeauthoryear{{Kriss}, {Hartig}, {Armus}, {Blair},
  {Caganoff}  \& {Dressel}}{{Kriss} et~al.}{1991}]{Kriss1991}
{Kriss} G.~A.,  {Hartig} G.~F.,  {Armus} L.,  {Blair} W.~P.,  {Caganoff} S.,
  {Dressel} L.,  1991, \mn@doi [\apjl] {10.1086/186105}, \href
  {http://adsabs.harvard.edu/abs/1991ApJ...377L..13K} {377, L13}

\bibitem[\protect\citeauthoryear{{Landt}, {Ward}, {Steenbrugge}  \&
  {Ferland}}{{Landt} et~al.}{2015a}]{Landt2015}
{Landt} H.,  {Ward} M.~J.,  {Steenbrugge} K.~C.,   {Ferland} G.~J.,  2015a,
  \mn@doi [\mnras] {10.1093/mnras/stv062}, \href
  {http://adsabs.harvard.edu/abs/2015MNRAS.449.3795L} {449, 3795}

\bibitem[\protect\citeauthoryear{{Landt}, {Ward}, {Steenbrugge}  \&
  {Ferland}}{{Landt} et~al.}{2015b}]{Landt2015b}
{Landt} H.,  {Ward} M.~J.,  {Steenbrugge} K.~C.,   {Ferland} G.~J.,  2015b,
  \mn@doi [\mnras] {10.1093/mnras/stv2176}, \href
  {http://adsabs.harvard.edu/abs/2015MNRAS.454.3688L} {454, 3688}

\bibitem[\protect\citeauthoryear{{Lightman} \& {Eardley}}{{Lightman} \&
  {Eardley}}{1974}]{Lightman1974}
{Lightman} A.~P.,  {Eardley} D.~M.,  1974, \mn@doi [\apjl] {10.1086/181377},
  \href {https://ui.adsabs.harvard.edu/abs/1974ApJ...187L...1L} {187, L1}

\bibitem[\protect\citeauthoryear{{Lipunov} et~al.,}{{Lipunov}
  et~al.}{2010}]{Lipunov2010}
{Lipunov} V.,  et~al., 2010, \mn@doi [Adv. Astron.] {10.1155/2010/349171},
  \href {http://adsabs.harvard.edu/abs/2010AdAst2010E..30L} {2010, 349171}

\bibitem[\protect\citeauthoryear{{Liu}, {Liu}, {Cheng}, {Qiao}  \&
  {Yuan}}{{Liu} et~al.}{2020}]{Liu2020}
{Liu} Z.,  {Liu} H.-Y.,  {Cheng} H.,  {Qiao} E.,   {Yuan} W.,  2020, \mn@doi
  [\mnras] {10.1093/mnras/stz3579}, \href
  {https://ui.adsabs.harvard.edu/abs/2020MNRAS.492.2335L} {492, 2335}

\bibitem[\protect\citeauthoryear{{Lyutyj}, {Oknyanskij}  \& {Chuvaev}}{{Lyutyj}
  et~al.}{1984}]{Lyuty1984}
{Lyutyj} V.~M.,  {Oknyanskij} V.~L.,   {Chuvaev} K.~K.,  1984, Soviet Astronomy
  Letters, \href {https://ui.adsabs.harvard.edu/abs/1984SvAL...10..335L} {10,
  335}

\bibitem[\protect\citeauthoryear{{MacLeod} et~al.,}{{MacLeod}
  et~al.}{2019}]{MacLeod2019}
{MacLeod} C.~L.,  et~al., 2019, \mn@doi [\apj] {10.3847/1538-4357/ab05e2},
  \href {https://ui.adsabs.harvard.edu/abs/2019ApJ...874....8M} {874, 8}

\bibitem[\protect\citeauthoryear{Marco \& Prieto}{Marco \&
  Prieto}{2005}]{Oliver2005}
Marco O.,  Prieto A.,  2005, in Brandner W.,  Kasper M.~E.,  eds, Science with
  Adaptive Optics. Springer Berlin Heidelberg, Berlin, Heidelberg, pp 315--320

\bibitem[\protect\citeauthoryear{{McHardy} et~al.,}{{McHardy}
  et~al.}{2014}]{Mchardy2014}
{McHardy} I.~M.,  et~al., 2014, \mn@doi [\mnras] {10.1093/mnras/stu1636}, \href
  {https://ui.adsabs.harvard.edu/abs/2014MNRAS.444.1469M} {444, 1469}

\bibitem[\protect\citeauthoryear{{Netzer}}{{Netzer}}{1974}]{Netzer1974}
{Netzer} H.,  1974, \mn@doi [\mnras] {10.1093/mnras/169.3.579}, \href
  {http://adsabs.harvard.edu/abs/1974MNRAS.169..579N} {169, 579}

\bibitem[\protect\citeauthoryear{{Netzer}}{{Netzer}}{2015}]{Netzer2015}
{Netzer} H.,  2015, \mn@doi [\araa] {10.1146/annurev-astro-082214-122302},
  \href {http://adsabs.harvard.edu/abs/2015ARA%26A..53..365N} {53, 365}

\bibitem[\protect\citeauthoryear{{Ochmann}, {Kollatschny}  \&
  {Zetzl}}{{Ochmann} et~al.}{2020}]{Kollatschny2020}
{Ochmann} M.~W.,  {Kollatschny} W.,   {Zetzl} M.,  2020, \mn@doi [Contributions
  of the Astronomical Observatory Skalnate Pleso]
  {10.31577/caosp.2020.50.1.318}, \href
  {https://ui.adsabs.harvard.edu/abs/2020CoSka..50..318O} {50, 318}

\bibitem[\protect\citeauthoryear{{Oknyanskii}, {Lyutyi}  \&
  {Chuvaev}}{{Oknyanskii} et~al.}{1991}]{Oknyansky1991}
{Oknyanskii} V.~L.,  {Lyutyi} V.~M.,   {Chuvaev} K.~K.,  1991, Soviet Astronomy
  Letters, \href {http://adsabs.harvard.edu/abs/1991SvAL...17..100O} {17, 100}

\bibitem[\protect\citeauthoryear{{Oknyanskij}}{{Oknyanskij}}{1988}]{Oknyansky1988}
{Oknyanskij} V.~L.,  1988, Peremennye Zvezdy, \href
  {https://ui.adsabs.harvard.edu/abs/1988PZ.....22..956O} {22, 956}

\bibitem[\protect\citeauthoryear{{Oknyanskij}}{{Oknyanskij}}{1989}]{Oknyansky1989}
{Oknyanskij} V.~L.,  1989, Soobshcheniya Spetsial'noj Astrofizicheskoj
  Observatorii, \href {https://ui.adsabs.harvard.edu/abs/1989SoSAO..61...70O}
  {61, 70}

\bibitem[\protect\citeauthoryear{{Oknyanskij} \& {Chuvaev}}{{Oknyanskij} \&
  {Chuvaev}}{1982}]{Oknyansky1982}
{Oknyanskij} V.~L.,  {Chuvaev} K.~K.,  1982, Astronomicheskij Tsirkulyar, \href
  {http://adsabs.harvard.edu/abs/1982ATsir1228....1O} {1228, 1}

\bibitem[\protect\citeauthoryear{{Oknyanskij} \& {Horne}}{{Oknyanskij} \&
  {Horne}}{2001}]{Oknyansky2001}
{Oknyanskij} V.~L.,  {Horne} K.,  2001, in {Peterson} B.~M.,  {Pogge} R.~W.,
  {Polidan} R.~S.,  eds,  Astronomical Society of the Pacific Conference Series
  Vol. 224, Probing the Physics of Active Galactic Nuclei. p.~149

\bibitem[\protect\citeauthoryear{{Oknyansky}, {Gaskell}  \&
  {Shimanovskaya}}{{Oknyansky} et~al.}{2015}]{Oknyansky2015}
{Oknyansky} V.~L.,  {Gaskell} C.~M.,   {Shimanovskaya} E.~V.,  2015, Odessa
  Astronomical Publications, \href
  {https://ui.adsabs.harvard.edu/abs/2015OAP....28..175O} {28, 175}

\bibitem[\protect\citeauthoryear{{Oknyansky} et~al.,}{{Oknyansky}
  et~al.}{2017}]{Oknyansky2017a}
{Oknyansky} V.~L.,  et~al., 2017, \mn@doi [\mnras] {10.1093/mnras/stx149},
  \href {http://adsabs.harvard.edu/abs/2017MNRAS.467.1496O} {467, 1496}

\bibitem[\protect\citeauthoryear{{Oknyansky}, {Lipunov}, {Gorbovskoy},
  {Winkler}, {van Wyk}, {Tsygankov}  \& {Buckley}}{{Oknyansky}
  et~al.}{2018}]{Oknyansky2018b}
{Oknyansky} V.~L.,  {Lipunov} V.~M.,  {Gorbovskoy} E.~S.,  {Winkler} H.,  {van
  Wyk} F.,  {Tsygankov} S.,   {Buckley} D.~A.~H.,  2018, The Astronomer's
  Telegram, \href {http://adsabs.harvard.edu/abs/2018ATel11915....1O} {11915}

\bibitem[\protect\citeauthoryear{{Oknyansky}, {Winkler}, {Tsygankov},
  {Lipunov}, {Gorbovskoy}, {van Wyk}, {Buckley}  \& {Tyurina}}{{Oknyansky}
  et~al.}{2019a}]{Oknyansky2019b}
{Oknyansky} V.~L.,  {Winkler} H.,  {Tsygankov} S.~S.,  {Lipunov} V.~M.,
  {Gorbovskoy} E.~S.,  {van Wyk} F.,  {Buckley} D.~A.~H.,   {Tyurina} N.~V.,
  2019a, \mn@doi [Odessa Astronomical Publications]
  {10.18524/1810-4215.2019.32.182514}, \href
  {https://ui.adsabs.harvard.edu/abs/2019OAP....32...75O} {32, 75}

\bibitem[\protect\citeauthoryear{{Oknyansky}, {Shenavrin}, {Metlova}  \&
  {Gaskell}}{{Oknyansky} et~al.}{2019b}]{Oknyansky2019a}
{Oknyansky} V.~L.,  {Shenavrin} V.~I.,  {Metlova} N.~V.,   {Gaskell} C.~M.,
  2019b, \mn@doi [Astronomy Letters] {10.1134/S1063773719040066}, \href
  {https://ui.adsabs.harvard.edu/abs/2019AstL...45..197O} {45, 197}

\bibitem[\protect\citeauthoryear{{Oknyansky}, {Winkler}, {Tsygankov},
  {Lipunov}, {Gorbovskoy}, {van Wyk}, {Buckley}  \& {Tyurina}}{{Oknyansky}
  et~al.}{2019c}]{Oknyansky2019}
{Oknyansky} V.~L.,  {Winkler} H.,  {Tsygankov} S.~S.,  {Lipunov} V.~M.,
  {Gorbovskoy} E.~S.,  {van Wyk} F.,  {Buckley} D.~A.~H.,   {Tyurina} N.~V.,
  2019c, \mn@doi [\mnras] {10.1093/mnras/sty3133}, \href
  {http://adsabs.harvard.edu/abs/2019MNRAS.483..558O} {483, 558}

\bibitem[\protect\citeauthoryear{{Osterbrock} \& {Shuder}}{{Osterbrock} \&
  {Shuder}}{1982}]{Osterbrock1982}
{Osterbrock} D.~E.,  {Shuder} J.~M.,  1982, \mn@doi [\apjs] {10.1086/190793},
  \href {http://adsabs.harvard.edu/abs/1982ApJS...49..149O} {49, 149}

\bibitem[\protect\citeauthoryear{{Parker} et~al.,}{{Parker}
  et~al.}{2016}]{Parker2016}
{Parker} M.~L.,  et~al., 2016, \mn@doi [\mnras] {10.1093/mnras/stw1449}, \href
  {http://adsabs.harvard.edu/abs/2016MNRAS.461.1927P} {461, 1927}

\bibitem[\protect\citeauthoryear{{Parker} et~al.,}{{Parker}
  et~al.}{2019}]{Parker2019}
{Parker} M.~L.,  et~al., 2019, \mn@doi [\mnras] {10.1093/mnrasl/sly224}, \href
  {http://adsabs.harvard.edu/abs/2019MNRAS.483L..88P} {483, L88}

\bibitem[\protect\citeauthoryear{{Pastoriza} \& {Gerola}}{{Pastoriza} \&
  {Gerola}}{1970}]{Pastoriza1970}
{Pastoriza} M.,  {Gerola} H.,  1970, \aplett, \href
  {http://adsabs.harvard.edu/abs/1970ApL.....6..155P} {6, 155}

\bibitem[\protect\citeauthoryear{{Penfold}}{{Penfold}}{1979}]{Penfold1979}
{Penfold} J.~E.,  1979, \mn@doi [\mnras] {10.1093/mnras/186.2.297}, \href
  {https://ui.adsabs.harvard.edu/abs/1979MNRAS.186..297P} {186, 297}

\bibitem[\protect\citeauthoryear{{Penston} \& {Perez}}{{Penston} \&
  {Perez}}{1984}]{Penston1984}
{Penston} M.~V.,  {Perez} E.,  1984, \mn@doi [\mnras]
  {10.1093/mnras/211.1.33P}, \href
  {https://ui.adsabs.harvard.edu/abs/1984MNRAS.211P..33P} {211, 33P}

\bibitem[\protect\citeauthoryear{{Penston}, {Fosbury}, {Boksenberg}, {Ward}  \&
  {Wilson}}{{Penston} et~al.}{1984}]{Penston1984a}
{Penston} M.~V.,  {Fosbury} R.~A.~E.,  {Boksenberg} A.,  {Ward} M.~J.,
  {Wilson} A.~S.,  1984, \mn@doi [\mnras] {10.1093/mnras/208.2.347}, \href
  {https://ui.adsabs.harvard.edu/abs/1984MNRAS.208..347P} {208, 347}

\bibitem[\protect\citeauthoryear{{Quintana}, {Kaufmann}  \& {Sasic}}{{Quintana}
  et~al.}{1975}]{Quintana1975}
{Quintana} H.,  {Kaufmann} P.,   {Sasic} J.~L.,  1975, \mn@doi [\mnras]
  {10.1093/mnras/173.1.57P}, \href
  {https://ui.adsabs.harvard.edu/abs/1975MNRAS.173P..57Q} {173, 57P}

\bibitem[\protect\citeauthoryear{{Rose}, {Elvis}  \& {Tadhunter}}{{Rose}
  et~al.}{2015}]{Rose2015}
{Rose} M.,  {Elvis} M.,   {Tadhunter} C.~N.,  2015, \mn@doi [\mnras]
  {10.1093/mnras/stv113}, \href
  {http://adsabs.harvard.edu/abs/2015MNRAS.448.2900R} {448, 2900}

\bibitem[\protect\citeauthoryear{{Ruan}, {Anderson}, {Eracleous}, {Green},
  {Haggard}, {MacLeod}, {Runnoe}  \& {Sobolewska}}{{Ruan}
  et~al.}{2019}]{Ruan2019}
{Ruan} J.~J.,  {Anderson} S.~F.,  {Eracleous} M.,  {Green} P.~J.,  {Haggard}
  D.,  {MacLeod} C.~L.,  {Runnoe} J.~C.,   {Sobolewska} M.~A.,  2019, arXiv
  e-prints, \href {https://ui.adsabs.harvard.edu/abs/2019arXiv190904676R} {p.
  arXiv:1909.04676}

\bibitem[\protect\citeauthoryear{{Rudy}, {Cohen}  \& {Ake}}{{Rudy}
  et~al.}{1988}]{Rudy1988}
{Rudy} R.~J.,  {Cohen} R.~D.,   {Ake} T.~B.,  1988, \mn@doi [\apj]
  {10.1086/166642}, \href
  {https://ui.adsabs.harvard.edu/abs/1988ApJ...332..172R} {332, 172}

\bibitem[\protect\citeauthoryear{{Runnoe} et~al.,}{{Runnoe}
  et~al.}{2016}]{Runnoe2016}
{Runnoe} J.~C.,  et~al., 2016, \mn@doi [\mnras] {10.1093/mnras/stv2385}, \href
  {https://ui.adsabs.harvard.edu/abs/2016MNRAS.455.1691R} {455, 1691}

\bibitem[\protect\citeauthoryear{{Shakura} \& {Sunyaev}}{{Shakura} \&
  {Sunyaev}}{1973}]{Shakura1973}
{Shakura} N.~I.,  {Sunyaev} R.~A.,  1973, \aap, \href
  {https://ui.adsabs.harvard.edu/abs/1973A&A....24..337S} {500, 33}

\bibitem[\protect\citeauthoryear{{Shapovalova} et~al.,}{{Shapovalova}
  et~al.}{2004}]{Shapovalova2004}
{Shapovalova} A.~I.,  et~al., 2004, \mn@doi [\aap]
  {10.1051/0004-6361:20035652}, \href
  {https://ui.adsabs.harvard.edu/abs/2004A&A...422..925S} {422, 925}

\bibitem[\protect\citeauthoryear{{Shapovalova} et~al.,}{{Shapovalova}
  et~al.}{2008}]{Shapovalova2008}
{Shapovalova} A.~I.,  et~al., 2008, \mn@doi [\aap]
  {10.1051/0004-6361:20079111}, \href
  {https://ui.adsabs.harvard.edu/abs/2008A&A...486...99S} {486, 99}

\bibitem[\protect\citeauthoryear{{Shapovalova} et~al.,}{{Shapovalova}
  et~al.}{2019}]{Shapovalova2019}
{Shapovalova} A.~I.,  et~al., 2019, \mn@doi [\mnras] {10.1093/mnras/stz692},
  \href {https://ui.adsabs.harvard.edu/abs/2019MNRAS.485.4790S} {485, 4790}

\bibitem[\protect\citeauthoryear{{Shappee} et~al.,}{{Shappee}
  et~al.}{2014}]{Shappee2014}
{Shappee} B.~J.,  et~al., 2014, \mn@doi [\apj] {10.1088/0004-637X/788/1/48},
  \href {http://adsabs.harvard.edu/abs/2014ApJ...788...48S} {788, 48}

\bibitem[\protect\citeauthoryear{{Sheng}, {Wang}, {Jiang}, {Yang}, {Yan}, {Dou}
   \& {Peng}}{{Sheng} et~al.}{2017}]{Sheng2017}
{Sheng} Z.,  {Wang} T.,  {Jiang} N.,  {Yang} C.,  {Yan} L.,  {Dou} L.,   {Peng}
  B.,  2017, \mn@doi [\apjl] {10.3847/2041-8213/aa85de}, \href
  {https://ui.adsabs.harvard.edu/abs/2017ApJ...846L...7S} {846, L7}

\bibitem[\protect\citeauthoryear{{Shobbrook}}{{Shobbrook}}{1966}]{Shobbrook1966}
{Shobbrook} R.~R.,  1966, \mn@doi [\mnras] {10.1093/mnras/131.3.365}, \href
  {http://adsabs.harvard.edu/abs/1966MNRAS.131..365S} {131, 365}

\bibitem[\protect\citeauthoryear{{{\'S}niegowska} \& {Czerny}}{{{\'S}niegowska}
  \& {Czerny}}{2019}]{Sniegowska2019}
{{\'S}niegowska} M.,  {Czerny} B.,  2019, arXiv e-prints, \href
  {https://ui.adsabs.harvard.edu/abs/2019arXiv190406767S} {p. arXiv:1904.06767}

\bibitem[\protect\citeauthoryear{{Storey} \& {Zeippen}}{{Storey} \&
  {Zeippen}}{2000}]{Storey2000}
{Storey} P.~J.,  {Zeippen} C.~J.,  2000, \mn@doi [\mnras]
  {10.1046/j.1365-8711.2000.03184.x}, \href
  {http://adsabs.harvard.edu/abs/2000MNRAS.312..813S} {312, 813}

\bibitem[\protect\citeauthoryear{{Vasudevan}, {Mushotzky}, {Winter}  \&
  {Fabian}}{{Vasudevan} et~al.}{2009}]{Vasudevan2009}
{Vasudevan} R.~V.,  {Mushotzky} R.~F.,  {Winter} L.~M.,   {Fabian} A.~C.,
  2009, \mn@doi [\mnras] {10.1111/j.1365-2966.2009.15371.x}, \href
  {https://ui.adsabs.harvard.edu/abs/2009MNRAS.399.1553V} {399, 1553}

\bibitem[\protect\citeauthoryear{{Veilleux}}{{Veilleux}}{1988}]{Veilleux1988}
{Veilleux} S.,  1988, \mn@doi [\aj] {10.1086/114766}, \href
  {http://adsabs.harvard.edu/abs/1988AJ.....95.1695V} {95, 1695}

\bibitem[\protect\citeauthoryear{{Winkler}}{{Winkler}}{1992}]{Winkler1992}
{Winkler} H.,  1992, \mn@doi [\mnras] {10.1093/mnras/257.4.677}, \href
  {http://adsabs.harvard.edu/abs/1992MNRAS.257..677W} {257, 677}

\bibitem[\protect\citeauthoryear{{Woo} \& {Urry}}{{Woo} \&
  {Urry}}{2002}]{Woo2002}
{Woo} J.-H.,  {Urry} C.~M.,  2002, \mn@doi [\apj] {10.1086/342878}, \href
  {https://ui.adsabs.harvard.edu/abs/2002ApJ...579..530W} {579, 530}

\bibitem[\protect\citeauthoryear{{da Silva}, {Steiner}  \& {Menezes}}{{da
  Silva} et~al.}{2017}]{Silva2017}
{da Silva} P.,  {Steiner} J.~E.,   {Menezes} R.~B.,  2017, \mn@doi [\mnras]
  {10.1093/mnras/stx1458}, \href
  {http://adsabs.harvard.edu/abs/2017MNRAS.470.3850D} {470, 3850}

\bibitem[\protect\citeauthoryear{{de Vaucouleurs}}{{de
  Vaucouleurs}}{1973}]{Vaucouleurs1973}
{de Vaucouleurs} G.,  1973, \mn@doi [\apj] {10.1086/152028}, \href
  {https://ui.adsabs.harvard.edu/abs/1973ApJ...181...31D} {181, 31}

\bibitem[\protect\citeauthoryear{{de Vaucouleurs} \& {de Vaucouleurs}}{{de
  Vaucouleurs} \& {de Vaucouleurs}}{1961}]{Vaucouleurs1961}
{de Vaucouleurs} G.,  {de Vaucouleurs} A.,  1961, \memras, \href
  {http://adsabs.harvard.edu/abs/1961MmRAS..68...69D} {68, 69}

\makeatother
\end{thebibliography}

\bsp	

\label{lastpage}
\end{document}